\documentclass[aps,prl,twocolumn,floatfix,superscriptaddress]{revtex4-1}
\usepackage{bm}
\usepackage{amssymb}
\usepackage{amsfonts}
\usepackage{graphicx}
\usepackage{dcolumn}
\usepackage{color}
\usepackage{setspace}
\usepackage{longtable}
\usepackage{hyperref}

\begin{document}

\title{Origin of the different conductive behavior in pentavalent-ion-doped anatase and rutile TiO$_2$}

\author{Kesong Yang}
\email{kesong.yang@gmail.com}
\altaffiliation{Present Address: Duke University, NC 27708, USA}
\affiliation{School of Physics, State Key Laboratory of Crystal Materials, Shandong University, Jinan 250100, China}
\affiliation{Department of Physics, National University of Singapore, Singapore 117542, Singapore}
\author{Ying Dai}
\email{daiy60@sina.com}
\affiliation{School of Physics, State Key Laboratory of Crystal Materials, Shandong University, Jinan 250100, China}
\author{Baibiao Huang}
\affiliation{School of Physics, State Key Laboratory of Crystal Materials, Shandong University, Jinan 250100, China}
\author{Yuan Ping Feng}
\email{phyfyp@nus.edu.sg}
\affiliation{Department of Physics, National University of Singapore, Singapore 117542, Singapore}

\begin{abstract}

The electronic properties of pentavalent-ion (Nb$^{5+}$, Ta$^{5+}$, and I$^{5+}$) doped anatase and rutile TiO$_2$ are studied using spin-polarized GGA+\emph{U} calculations. Our calculated results indicate that these two phases of TiO$_2$ exhibit different conductive behavior upon doping. For doped anatase TiO$_2$, some up-spin-polarized Ti 3\emph{d} states lie near the conduction band bottom and cross the Fermi level, showing an \emph{n}-type half-metallic character. For doped rutile TiO$_2$, the Fermi level is pinned between two up-spin-polarized Ti 3\emph{d} gap states, showing an insulating character. These results can account well for the experimental different electronic transport properties in Nb (Ta)-doped anatase and rutile TiO$_2$.
\end{abstract}
\maketitle

Transparent conducting oxides (TCOs) have many applications in optoelectronic devices such as flat panel displays, organic light-emitting diodes and solar cells. As one of the TCOs, Sn-doped In$_2$O$_3$ is widely used because of its excellent optical transparency and electrical transport property.\cite{Edwards2004_DaltonTran} However, owning to the high cost of indium and the increasing demand for high-performance TOCs, many efforts have been made to develop new TOCs materials.\cite{Tadatsugu2005SST, Wang2010JAP} Recently, as one potential candidate of TCOs, Nb (Ta)-doped anatase TiO$_2$ has attracted lots of attention because of its high electrical conductivity and optical transparency.\cite{Furubayashi2005APL, Hitosugi2005JJAP,Furubayashi2005TSF,Furubayashi2007JAP,Gillispie2007JAP, Gillispie2007JMR,Hoang2008APE,Archana2011APL} However, the origin of its high conductivity is still controversial. Wan \emph{et al.}\cite{Wan2006APL} found that Nb-doped TiO$_2$ grown on (0001) Al$_2$O$_3$ substrate shows a much larger resistivity than that grown on (100) SrTiO$_3$ substrate, and hence the Nb diffusion into the SrTiO$_3$ substrate is thought to lead to the high conductivity of Nb-doped TiO$_2$. Meanwhile, Furubayashi \emph{et al.}\cite{Furubayashi2006APL} confirmed that Nb-doped TiO$_2$ forms a rutile phase on the (0001) Al$_2$O$_3$ substrate but an anatase phase on the other substrates, and hence they suggested that the higher resistivity of Nb-doped TiO$_2$ grown on (0001) Al$_2$O$_3$ substrates is caused by the formation of rutile phase. Interestingly, later experiments further verified that Nb (Ta)-doped anatase TiO$_2$ is metallic but Nb (Ta)-doped rutile TiO$_2$ is insulating.\cite{Zhang2007JAP,Barman2011APL} Therefore, one may speculate that the Nb (Ta) doping can lead to different conductive properties in anatase and rutile TiO$_2$, i.e., conductive for anatase phase but insulating for rutile phase.\cite{Yang2009ICMAT}

In principle, a pentavalent dopant, such as Nb$^{5+}$, Ta$^{5+}$, and I$^{5+}$,\cite{Yang_2008_CM,Yang2012REVIEW} can release one additional electron into TiO$_2$ than a Ti$^{4+}$, and introduce donor levels.\cite{Finazzi_2008_JCP,Hitosugi2008APE} Although Nb-doped anatase TiO$_2$ has been studied using standard density functional theory (DFT) calculations,\cite{Hitosugi2008APE, Liu2008APL, Kamisaka2009JCP} the standard DFT calculations within either local density approximation (LDA) or generalized gradient approximation (GGA) cannot properly deal with the strong-corrected Ti 3\emph{d} electrons.\cite{Finazzi_2008_JCP,Yang_2010_PRB_TiO} In the present work, we studied the electronic properties of pentavalent-ion (Nb$^{5+}$, Ta$^{5+}$, and I$^{5+}$) doped anatase and rutile TiO$_2$, respectively, using spin-polarized GGA+\emph{U} calculations which can give a proper description of Ti 3\emph{d} orbitals.\cite{Finazzi_2008_JCP,Yang_2010_PRB_TiO,Yang2009CPC}. Our calculations indicate that Nb (Ta, I)-doped anatase TiO$_2$ shows an \emph{n}-type half-metallic character, while Nb (Ta, I)-doped rutile TiO$_2$ shows an insulating character. These results give a good explanation for experimentally observed different conductive behavior in Nb (Ta)-doped TiO$_2$.

The spin-polarized GGA+\emph{U} electronic structure calculations were carried out using the Vienna \emph{ab-inito} simulation package (VASP).\cite{VASP_PRB, VASP_CMS} 108-atom $3\times 3\times 3$ supercell of the anatase phase and 72-atom $2\times 2\times 3$ supercell of the rutile phase are used to model Nb (Ta, I)-doped TiO$_2$, in which a Ti atom is substituted by a Nb (Ta, I) atom. The projector augmented wave (PAW) potentials are used to treat electron-ion interactions and generalized gradient approximation parameterized by Perdew and Wang (PW91) are used for electron exchange-correction functional.\cite{PAW, PW91} A cut-off energy of 500 eV and a $3\times 3\times 3$ \emph{k}-point mesh centered at $\Gamma$ point are used. The lattice parameters and all the atomic positions are fully relaxed until all components of the residual forces are smaller than 0.01 eV/\AA. In our GGA+\emph{U} calculations, the on-site effective \emph{U} parameter (\emph{U}$_{eff}$=\emph{U}-\emph{J}=5.8 eV) proposed by Dudarev \emph{et al}. is adopted for Ti 3\emph{d} electrons.\cite{Dudarev1998PRB,Anisimov1991PRB}

\begin{figure}[ht]
\center
\includegraphics[scale=0.4]{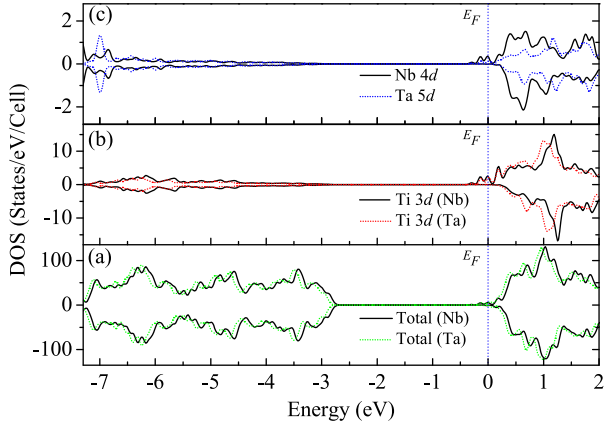}
\caption{(Color online) Calculated (a) TDOS and PDOS plots for (b) Ti 3\emph{d} and (c) Nb 4\emph{d} (Ta 5\emph{d} ) states for Nb (Ta)-doped anatase TiO$_2$. The Fermi level is indicated by the vertical dot line at 0 eV.
}\label{f1}
\end{figure}

\begin{figure}[ht]
\center
\includegraphics[scale=0.4]{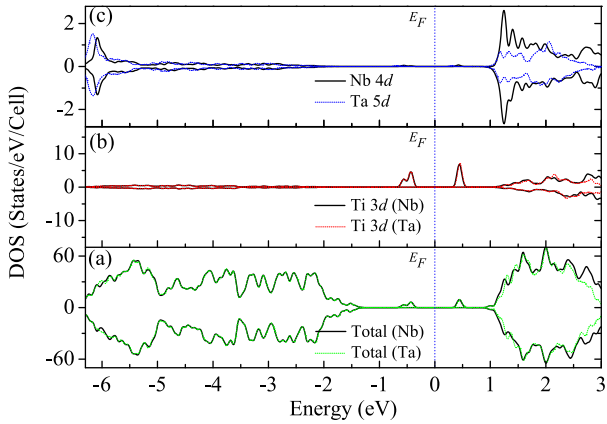}
\caption{(Color online) Calculated (a) TDOS and PDOS plots for (b) Ti 3\emph{d} and (c) Nb 4\emph{d} (Ta 5\emph{d} ) states for Nb (Ta)-doped rutile TiO$_2$. The Fermi level is indicated by the vertical dot line at 0 eV.
}\label{f2}
\end{figure}

\begin{figure}[ht]
\center
\includegraphics[scale=0.4]{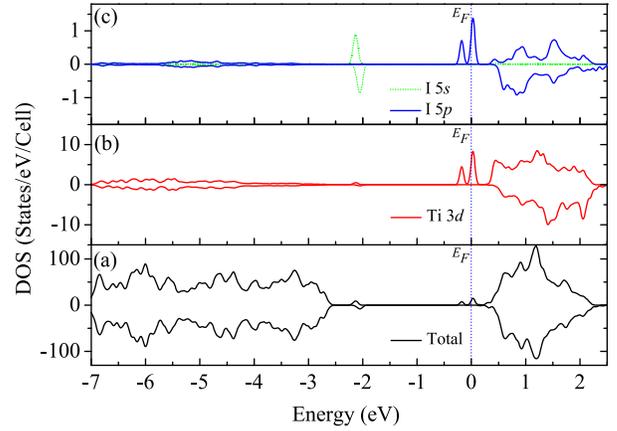}
\caption{(Color online) Calculated (a) TDOS and PDOS plots for (b) Ti 3\emph{d} and (c) I 5\emph{s}/5\emph{p} states for I-doped anatase TiO$_2$. The Fermi level is indicated by the vertical dot line at 0 eV.
}\label{f3}
\end{figure}

\begin{figure}[ht]
\center
\includegraphics[scale=0.4]{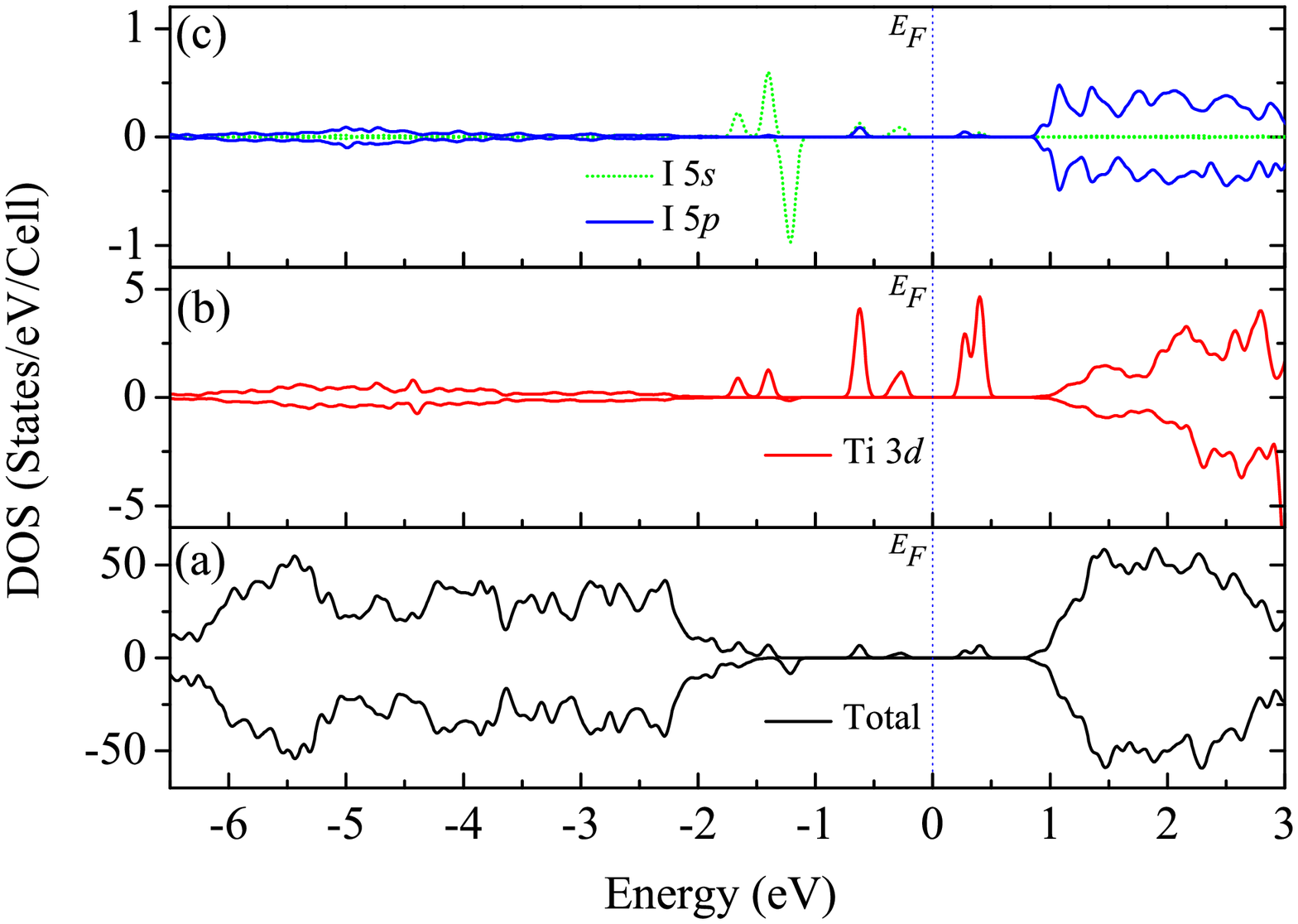}
\caption{(Color online) Calculated (a) TDOS and PDOS plots for (b) Ti 3\emph{d} and (c) I 5\emph{s}/5\emph{p} states for I-doped rutile TiO$_2$. The Fermi level is indicated by the vertical dot line at 0 eV.
}\label{f4}
\end{figure}

To examine the substitutional Nb (Ta, I) doping effects on the electronic property of TiO$_2$, we calculated the total density of states (TDOS) and partial density of states (PDOS) for Nb (Ta)-doped anatase (see Fig \ref{f1}) and rutile (see Fig \ref {f2}), and the TDOS and PDOS for I-doped anatase (see Fig \ref{f3}) and rutile (see Fig \ref{f4}).
For Nb (Ta)-doped anatase TiO$_2$, the calculated TDOS shows that it is spin-polarized, and some up-spin-polarized gap states extend from the conduction band (CB) into the band gap. These up-spin-polarized gap states are located just below the CB bottom, and cross the Fermi level, indicating an \emph{n}-type half-metallic character. This is in good agreement with the experimentally observed excellent conductive property in Nb (Ta)-doped anatase TiO$_2$.\cite{Furubayashi2005APL, Hitosugi2005JJAP,Furubayashi2005TSF,Furubayashi2007JAP,Gillispie2007JAP, Gillispie2007JMR,Hoang2008APE,Archana2011APL} The PDOS shows the Nb 4\emph{d}  (Ta 5\emph{d} ) states mix with the Ti 3\emph{d} states in the whole CB, indicating that the Nb (Ta) dopant forms a strong Nb (Ta)-O bond. It is also noted that the up-spin Ti 3\emph{d} orbitals strongly hybridize with the Nb 4\emph{d} (Ta 5\emph{d} ) orbitals, and as will be further discussed below, the Ti 3\emph{d} orbitals mostly contribute to the up-spin gap states. For Nb (Ta)-doped rutile phase (see Fig. \ref{f2}), as in the case of doped anatase, Nb 4\emph{d} (Ta 5\emph{d} ) states spread over the whole CB, and two fully spin-polarized gap states with a gap about 1.1 eV lie in the band gap. Its PDOS shows that these two gap states are also mainly contributed by the Ti 3\emph{d} states, however, the Fermi level lies between the two gap states, showing an insulating character. These calculated results can account for the experimentally observed much higher resistivity in Nb (Ta)-doped rutile phase than the anatase phase.\cite{Wan2006APL,Furubayashi2006APL,Zhang2007JAP,Barman2011APL}

\begin{figure*}[htbp]
\center
\includegraphics[scale=1.1]{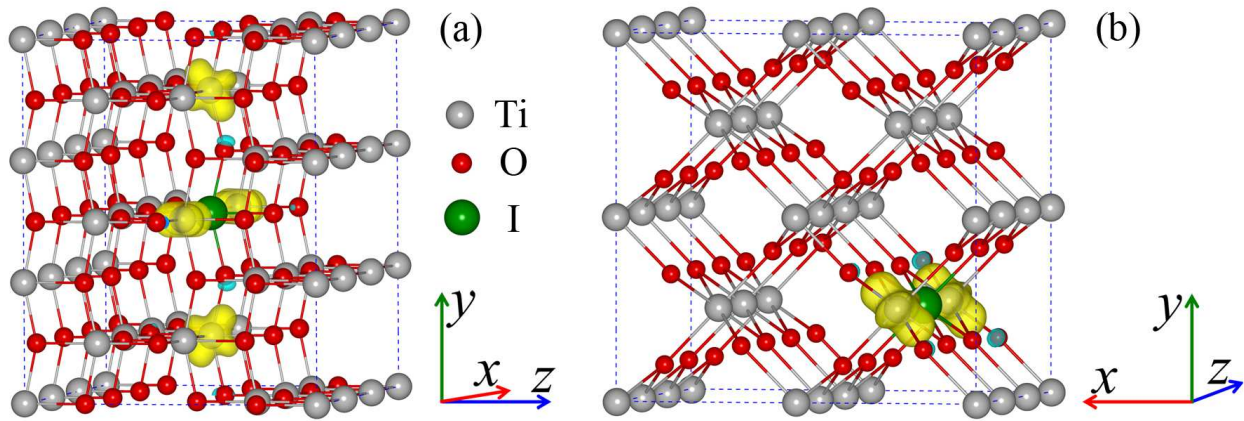}
\caption{(Color online) Calculated spin density distribution of I-doped anatase and rutile TiO$_2$.}
\label{f5}
\end{figure*}

In addition, it is worth mentioning that our calculated results are different from previous GGA+\emph{U} calculations done by Morgan \emph{et al.},\cite{Morgan2009JMC} in which the Nb (Ta)-doped anatase and rutile TiO$_2$ both show insulating properties. This discrepancy may be attributed to the following two reasons: (1) Morgan \emph{et al.} applied a \emph{U} parameter of 4.2 eV for Ti 3\emph{d} electrons, which is much smaller than the used value of 5.8 eV in this work. (2) In our GGA+\emph{U} calculations, all the degrees of the freedom for Nb (Ta)-doped TiO$_2$, including lattice parameters and all the atomic positions, are fully relaxed. In contrast, in Morgan \emph{et al.}'s calculations, only the internal degrees of freedom are allowed to relax. Therefore, the difference in \emph{U} values and structural optimization methods may be responsible for the different electronic properties of Nb (Ta)-doped TiO$_2$ in our calculations from those of Margan \emph{et al.} In fact, Orita also found a metallic character in Nb-doped anatase TiO$_2$ using GGA+\emph{U} calculations,\cite{Orita2010JJAP} in which all the atomic coordinates and lattice constants are optimized. In particular, recent hybrid density functional calculations also show that Nb (Ta)-doped anatase TiO$_2$ is metallic, while Nb (Ta)-doped rutile TiO$_2$ is semiconducting.\cite{Yamamoto2012PRB} In summary, these related studies further confirm our GGA+\emph{U} calculations. As a consequence, we can conclude that the conducting character of Nb (Ta)-doped anatase and the insulating character of Nb (Ta)-doped rutile are their intrinsic properties.

For I-doped anatase and rutile TiO$_2$, similar to the case of Nb (Ta) doping, an \emph{n}-type half-metallic character occurs in I-doped anatase phase but an insulating character occurs in I-doped rutile phase. The calculated TDOS and PDOS are shown in Fig \ref{f3} for anatase phase and Fig \ref{f4} for rutile phase. However, different from that of Nb (Ta) doping, a double filled gap state mostly consisting of I 5\emph{s} orbital appears in the band gap, which is located just above the valence band maximum. This indicates that I dopant exists as I$^{5+}$ (5\emph{s}$^2$5\emph{p}$^0$) in TiO$_2$, which is consistent with the standard GGA calculations.\cite{Yang_2008_CM}

To understand the origin of the different conductive behavior associated with the spin-polarized Ti 3\emph{d} states in Nb (Ta, I)-doped anatase and rutile TiO$_2$, we take I-doped TiO$_2$ as an example to show its three-dimensional spin density distribution in Fig. \ref{f5}. For I-doped anatase TiO$_2$, its spin density mostly comes from the four equivalent second-nearest Ti ions around the I dopant (see in Fig. \ref{f5}a). These four equivalent Ti ions share one electron donated by one I$^{5+}$ ion with their Ti 3\emph{d} orbitals, and thus produce a total spin magnetic moment of 1.0 $\mu_B$. Therefore, to a first approximation, these four equivalent Ti ions should exist as Ti$^{+3.75}$ (\emph{d}$^{1/4}$). Actually, experimental core-level photoemission spectra measurements showed a minor peak of the binding energy below that of Ti$^{4+}$ ion,\cite{Hitosugi2008APE} and this chemical shift corresponds to an increase of the valence electron density on Ti 3\emph{d} orbitals. For I-doped rutile TiO$_2$, in contrast, its spin density is mainly contributed by the two equivalent second-nearest Ti ions (see Fig \ref{f5}b). These two Ti ions share one electron and produce a total spin magnetic moment of 1.0 $\mu_B$, and thus one can assume that the two equivalent Ti ions exist as Ti$^{+3.5}$ (\emph{d}$^{1/2}$). As a result, it is expected that a lower binding energy of Ti$^{+3.5}$ ions than that of Ti$^{+3.75}$ ions can be observed in Nb (Ta, I)-doped rutile TiO$_2$ through experimental core-level photoemission spectra measurements. Furthermore, the increasing of the electron density on Ti 3\emph{d} orbitals directly leads to the splitting between Ti 3\emph{d} occupied states and unoccupied states, which is responsible for the insulating character of I-doped rutile TiO$_2$. Similar spin density distributions also occur in Nb (Ta)-doped anatase and rutile TiO$_2$.

In summary, we studied the electronic properties of Nb (Ta, I)-doped anatase and rutile TiO$_2$ by spin-polarized GGA+\emph{U} calculations. In doped anatase TiO$_2$, the Fermi level is pinned in some up-spin-polarized gap states near the CBM, showing an \emph{n}-type half-metallic conductive property. In doped rutile, in contrast, two localized states with a gap about 1.1 eV are introduced in the band gap, and the Fermi level lies between them, showing an insulating character. Therefore, to prepare the TiO$_2$-based TCOs through pentavalent-ion-doping, it is essential to avoid the phase transition from anatase to rutile. Our theoretical calculations may provide some useful guidance to develop TiO$_2$-based TCOs.

\begin{acknowledgments}
This work is supported by the National Basic Research Program of China (973 program, 2007CB613302), National Science foundation of China under Grant 11174180 and 20973102, and the Natural Science Foundation of Shandong Province under Grant number ZR2011AM009. Y.P.F is thankful for the support of the Singapore National Research Foundation Competitive Research Program (Grant No. NRF-G-CRP 2007-05).
\end{acknowledgments}

%\bibliography{TiO}

\end{document}